\documentclass[final,5p,times,twocolumn]{elsarticle}

\usepackage{subfigure}
\usepackage{graphicx}
\usepackage{epstopdf}
\usepackage{longtable}
\usepackage{amsmath}
\usepackage[load-configurations = abbreviations]{siunitx}
\usepackage{color}
\usepackage[usenames,dvipsnames,svgnames,table]{xcolor}
\usepackage{lineno,hyperref}
\usepackage{siunitx}

\modulolinenumbers[5]

\begin{document}

\author[LNF]{A. Marocchino \corref{mycorrespondingauthor} }
\cortext[correspondingauthor]{Corresponding author}
\ead{alberto.marocchino@lnf.infn.it,alberto.marocchino@gmail.com}
\author[LNF]{E. Chiadroni}
\author[LNF]{M. Ferrario}
\author[LNF]{F. Mira}
\author[INFN_MI]{A. R. Rossi}

\address[LNF]{Laboratori Nazionali di Frascati, INFN, Via E. Fermi 40, Frascati, Italy}
\address[INFN_MI]{INFN Milano, via Celoria 16, 20133 Milan, Italy}

\title{Design of high brightness Plasma Wakefield Acceleration experiment at SPARC\_LAB test facility with particle-in-cell simulations}

\date{\today}

\begin{abstract}
The present numerical investigation of a Plasma Wakefield Acceleration scenario in the weakly non linear regime with external injection is motivated by the upcoming campaigns at the SPARC\_LAB test facility where the final goal is to demonstrate modest gradient acceleration ($\sim$1 GV/m) with no quality loss. The accelerated bunch can be envisioned to seed a free electron laser. The numerical study has been conducted with the particle-in-cell code ${\tt ALaDyn}$, an exhaustive description of the plasma-acceleration version is provided. The configuration consider a two bunches setup with parameters in the facility range, the bunches are generated and pre-accelerated up to 100 MeV by a high brightness photo-injector prior plasma injection. To verify the working point robustness we have considered case scenario where the driver bunch reaches the plasma or with a larger dimension or with large emittance. We also present an analytical approach based on the envelope equation that allows to reduce the matching condition in the presence of a ramp. Here, we limit our interest to a simplified theoretical case with a linear plasma ramp. As a final aspect we propose to combine classical integrated bunch diagnostics with the test by Shapiro-Wilk, a mathematical test to diagnose bunch deviation from a Gaussian distribution.
\end{abstract}

\maketitle

\section{Introduction}
Plasma-based accelerators represent a new frontier for the development of compact advanced radiation sources and next generation linear colliders. High brightness electron beams are the future goal of such kind of particle accelerators in order to compete with those based on conventional RF photo-injectors. In the last decades great progresses have been achieved in several international laboratories \cite{Corde2015, Litos2014, Leemans2006,Schwoerer:2006fc,Blumenfeld:2007ja,Bingham:2006jh,Muggli2008,Blumenfeld2007,kallos2007,Muggli2004} to demonstrate the acceleration of electron beams with gradients of the order of several tens of GV/m, produced either by laser-driven (LWFA) or particle-driven plasma wakefields (PWFA). In PWFA the high gradient wakefield is induced by a high-energy charged particle beam (referred to as driver bunch) travelling through a pre-ionised plasma. The background electrons by shielding the charge breakup produced by the driver induce an accelerating field. The second bunch (referred to as trailing bunch or witness) placed on the right phase is so accelerated by the induced electric field \cite{vanderMeer1985,Katsouleas:1986,Rosenzweig1991,Litos:2014yqa}. 

In the present manuscript we report on PWFA numerical simulations whose final goal is to demonstrate trailing bunch modest acceleration ($\sim$1 GV/m) with no quality loss. The study is motivated by the upcoming SPARC\_LAB proof of principle experimental campaign. The considered setup assumes both driver and trailing bunch externally injected. Experimentally, the bunches are contemporarily extracted from a single photo-injector by mean of a laser modulation, the COMB technique in velocity bunching configuration \cite{Pompili2016}. The numerical investigation, in this paper, is limited to the plasma channel and it is conducted with the state-of-art particle-in-cell (PIC) code ${\tt ALaDyn}$ \cite{aladynDOI,Londrillo2010,Londrillo2014,Benedetti2008}. By recalling the envelope equation and the generalised PWFA transverse matching condition, we specialise to find out the required bunch dimensions and characteristics, such as emittance and energy spread, for our future experiments. To verify how sensitive is the parameters choice to the quality of acceleration we have degraded the driver quality by injecting it our of matching. The out-of-matching condition consists in drivers with a larger transverse dimension or with larger emittance. The varied parameters are just two since PIC simulations are computationally expensive and cannot permit for a systematic approach. Realistic case scenario should also consider some ramped density profile. 
Plasma ramps can relax matching condition allowing for injections at larger dimensions. To illustrate how to leverage on ramp beneficial effects a simple analytical formulation is reported. To generalise and simplify as much as possible the scenario we have considered a linear plasma ramp (reasonable assumption in the limit of a gas-nozzle profile, too simplified in the case of a capillary discharge \cite{Marocchino2017}).  In conclusion we also report on the Shapiro-Wilk  \cite{Royston1982} statistical test. The test is used, in addition to classical bunch quality indicator such as emittance and energy spread, to diagnose bunch deviation from a Gaussian distribution. Emittance and energy spread are integrated indicator that do not keep into account for the bunch shape. The Shapiro-Wilk test is used to verify that the trailing bunch retains its gaussian distribution over distance, while the driver does not.

\section{The PIC Code ${\tt ALaDyn}$}
Plasma wakefield acceleration scenarios presented in this paper have been numerically investigated with the use of the full 3D PIC code ${\tt ALaDyn}$, a code originally developed for LWFA \cite{Londrillo2014} and recently adapted for PWFA. In this section we briefly present the main code structure, the algorithm used to initialise the bunch self-consistent fields, the use of Twiss-parameters to manipulate the initial bunch shape and convergence, and we also present our implementation of the ADK ionization model. The ADK ionization implementation is recalled in this paper to keep the ${\tt ALaDyn}$ description as a whole despite this module is only used by the same authors, in this conference series and specifically in \cite{mira2017}, to investigate PWFA ionization injection. It is worth mentioning that ${\tt ALaDyn}$ has been supported for preliminary runs with the faster hybrid code Architect \cite{MarocchinoZenodo,Marocchino2016,Massimo2016}. 

\subsection{The Vlasov-Maxwell integration loops}
The reference model for relativistic fully kinetic PIC codes is given by the Vlasov-Maxwell system where plasma particles, as discrete characteristics of the Vlasov equation, are advanced in time by Lagrangian equation of motion; whereas self-consistent electromagnetic fields are evolved by the Maxwell equations.
In beam-driven PWFA regimes bunch and background plasma have different properties and cover quite separated phase-space domains. Moreover the injected bunches form a charged system, while the background plasma is globally a neutral system. It then appears natural to describe the whole phase space with two species: the background electrons (subscript $e$) and the bunch electrons (subscript $b$). Since ions can be considered static their effect is taken into account directly in the Maxwell's equations. If $(\mathbf{p}_s,\mathbf{x}_s)$, with $s=e,b$, denotes the particle phase space coordinates and $q_s$ the particle charge, for each component the relativistic equations of motion for our two-species system are,
\begin{eqnarray}
d_t \mathbf{p}_s &=&  q_s \left( \mathbf{E} + \frac{\mathbf{v}_s}{c} \times \mathbf{B} \right) \nonumber \\
d_t \mathbf{x}_s &=& \mathbf{v}_s
\end{eqnarray}
where $\mathbf{E}$ and $\mathbf{B}$ are the total electric and magnetic field evaluated at particle position.

Maxwell's equation for the total $\left( \mathbf{E}, \mathbf{B} \right)$ fields in Gaussian units are given in the usual Eulerian form,
\begin{eqnarray}
\nabla \cdot \mathbf{E} &=& 4 \pi \rho \nonumber \\
\nabla \cdot \mathbf{B} &=& 0 \nonumber \\
\partial_t \mathbf{B} &=& - c \: \nabla \times \mathbf{E} \nonumber \\
\partial_t \mathbf{E} &=&  c \:  \nabla \times \mathbf{B} - 4 \pi \mathbf{J}
\end{eqnarray}
where $\mathbf{J}$ and $\rho$ are,m the total current and the total charge density, respectively; and they are calculated as the superposition of bunch-plasma species contributions: $\rho=\rho_e+\rho_b$ and $\mathbf{J}=\mathbf{J}_e+\mathbf{J}_b$, respectively.

The electric and magnetic fields are solved with a standard second order Leap-Frog scheme on a staggered space-time grid (the Yee-lattice \cite{yee66}). To take into account the specific structure of a bunch-plasma system, where bunch charge-density and fields have a central role, we split the electro-magnetic fields in a rigidly advected part, fast moving component, and a residual, slowly moving part,
\begin{equation}
\mathbf{E}=\mathbf{E}^{0}+\mathbf{E}^{\prime},\quad
\mathbf{B}=\mathbf{B}^{0}+\mathbf{B}^{\prime},
\end{equation}
where $\mathbf{E}^{0}$ and $\mathbf{B}^{0}$ represent the advected solution, namely the exact static solution in a comoving coordinate system $\zeta = z- v_b t$ (with $z$ longitudinal propagating direction, and $v_b$ bunch velocity): $(\partial_t+v_b\partial_z) \mathbf{E}^0=0$ and $(\partial_t+v_b\partial_z)\mathbf{B}^0=0$; with initial conditions given by the initial bunch fields $ \mathbf{E}^0(t=0)=\mathbf{E}_b(t=0)$ and $\mathbf{B}^0(t=0)=\mathbf{B}_b(t=0)$. The same approach is also used for charge density ($\rho=\rho^{0}+\rho^{\prime}$). In this approach the rigidly advected parts $(\mathbf{E}^0,\,\mathbf{B}^0)$ are solution of the Maxwell equation with source terms $\rho^0$ and $J_z^0=v_b\rho^0$, whereas the residual part $(\mathbf{E}^{\prime},\,\mathbf{B}^{\prime},\,\rho^{\prime})$, starting from zero initial conditions $\mathbf{E}^{\prime}=\mathbf{B}^{\prime}=\rho^{\prime}=0$, satisfies the Maxwell's equations with source terms $(\rho-\rho^0)$ and $(J_z-J_z^0, \mathbf{J}_{\perp})$.

\subsection{Bunch self-consistent field initialisation}
Electron bunches are initialised in vacuum with a bi-gaussian (or normal-multivariate) charge distribution. We initialise bunches in vacuum to treat, self consistently, the transition from vacuum to plasma. The self-consistent fields are evaluated by a quasi-static approximation based on bunch density $\rho_b(t=0)$ and longitudinal current $J_z(t=0)=v_b\rho_b$ with negligible transverse current $\mathbf{J}_{\perp}=0$. For relativistic energies, i.e. large $\gamma$, and low emittance initial distributions, this approximation is clearly preserved on short advection time as a static solution in the comoving coordinate system. In terms of scalar potential $\varphi$, the Gauss law for a quasi-static approximation reduces to the Poisson-like elliptic equation
\begin{equation} \label{eqn:potential0}
- \left[\frac{1}{\gamma^2} \frac{\partial^2}{\partial z^2} + \frac{\partial^2}{\partial x^2} + \frac{\partial^2}{\partial y^2} \right] \varphi = \rho_b(t=0),
\end{equation}
where $\gamma^2=[1-\beta^2]^{-1}$ and $\beta=v_b/c$. Under the cold fluid assumption, $\mathbf{A}_{\perp}=0$, the Lorentz gauge condition gives $A_z=v_b\varphi$. The initial bunch fields can then be evaluated in terms of the $\varphi$ potential only,
\begin{eqnarray}
E_{b,z}=-\gamma^{-2} \partial_z \varphi, & \quad & B_{b,z}= 0, \nonumber \\
E_{b,x}=- \partial_x \varphi, & \quad & B_{b,x}= v_b E_{b,y}, \nonumber \\
E_{b,y}=- \partial_y \varphi, & \quad & B_{b,y}=- v_b E_{b,x}.
\end{eqnarray}
To solve for the initial potential $\varphi$ in free space a cosine transform is implemented using standard FFT algorithms.

\subsection{Bunch profile manipulation via phase space rotation}
${\tt ALaDyn}$ allow phase space manipulation to initialise bunches with arbitrary convergence, the manipulation is operated by a phase-space rotation via Twiss parameters \cite{Reiser,Chao}. Conventionally electron bunches are initialised at waist, Twiss-$\alpha$:equal zero. Bunches at waist represent a due simplification to study the underlying physics, indeed for more realistic case we need to include some degree of divergence, of bunch manipulation. We need to generate bunches that can either be converging or diverging, preserving emittance and energy spread during such a rotation, so to comfortably generate cases that would reach the waist after a chosen travelled distance. 

To transform a bunch from the waist condition to the desired configuration, we apply a phase-space linear transformation for each single particle, 
\begin{equation}
\begin{bmatrix}
    x_{\rm new} \\
    p_{x - {\rm new}}
\end{bmatrix}
=
\begin{bmatrix}
    s_{11} & s_{12} \\
    0 & s^{-1}_{11}
\end{bmatrix}
\begin{bmatrix}
    x \\
    p_x
\end{bmatrix},
\end{equation}
where $x$ and $p_x$ are the original -at waist- particle phase-space coordinate, while $(x_{\rm new}, p_{x - {\rm new}})$ is the new coordinate. The transformation matrix is constructed with a determinant equal to 1 to preserve emittance over transformation. The matrix elements functions of Twiss parameters $\alpha_T$ and $\beta_T$ have the following expression,
\begin{eqnarray} \label{twissrotation}
s_{11}&=&\sqrt{ \frac{\varepsilon_{x,rms} \: \beta_{T}} {\sigma^2_x (1+\alpha^2_{T})} }, \nonumber \\
s_{12}&=&-\frac{s_{11} \alpha_{T} \sigma^2_x}{\varepsilon_{x,rms}},
\end{eqnarray}
with  $\varepsilon_{x,rms}$ the transverse normalised rms-emittance, $\alpha_{T}=-\sigma_{xp_x-{\rm new}}/\varepsilon_{x,rms}$ and $\beta_{T}=\sigma^2_{x-{\rm new}}/\varepsilon_{x,rms}$. $\sigma_{x-{\rm new}}$ is the transformed rms-trasverse dimension ($\sigma_{x-{\rm new}}=\langle x^2_{\rm new} \rangle )$, $\sigma_{xp_x-{\rm new}}$ the transformed cross-correlation ($\sigma_{xp_x-{\rm new}}= \langle x_{\rm new} \: p_{x-\rm new}\rangle$). The same procedure is also applied to $y$ direction. In section 3 we will show how to combine Eq.(\ref{twissrotation}) with the envelope equation to generate the desired bunch.

\subsection{Ionization Models}
The latest ${\tt ALaDyn}$ releases also includes an ionisation module, that could be chosen between the Ammosov-Delone-Krainov (ADK) \cite{ADK} model or the Barrier-Suppression-Ionization (BSI) \cite{BSI} model. The ADK model is, at present, the generally considered and used model for PWFA problem. The choice is related to the intensity of the induced wakefields. The ionisation ADK rate considered is,
\begin{eqnarray} \label{eqn:adk}
& &W_{_{ADK}}(l) \: [s^{-1}] = \omega_{a} \: C_{n^*l^*}^2 (2l+1) \: \left(\frac{U_{i}}{2U_H}\right) \nonumber \\
& & \left(2\frac{E_a}{E}\left(\frac{U_i}{U_H}\right)^{3/2}\right)^{2n^*-1} 
\exp{\left(-\frac{2}{3}\frac{E_a}{E}\left(\frac{U_i}{U_H}\right)^{3/2}\right)}
\end{eqnarray} 
where $l$ is the electron's orbital quantum number, $\omega_a=\alpha^3c/r_e=$ \SI{4.13e16}{\per\second} is the atomic unit frequency, $U_i$ is the ionization potential of th $i^{th}$ level, and $U_H=13.6$ eV is a the ionization potential of Hydrogen at the fundamental state. The effective principal quantum number is defined as $n^*=Z\sqrt{U_H/U_i}$ with $Z$ the atom's atomic number,  $l^*$ is the effective value of the orbital number and it is defined as $l^*=n^*_0-1$ with $n^*_0$ the effective principal quantum number of the ground state. Original ADK formulation also depends upon the quantum number $m$, the projection of $l$, in our calculation we assume that all bounded electrons are in their fundamental magnetic quantum number state characterised by $m=0$. Part of the computational cost for the evaluation of Eq.(\ref{eqn:adk}) is strictly connected with the evaluation of the coefficient $C_{n^*l^*}^2$ that in its original version \cite{ADK} retains two special $\Gamma$-functions. The computational cost is greatly reduced assuming an asymptotic behaviour for $C_{n^*l^*}^2$ in the limit of $l^* \ll n^*$, that reduces the coefficient to the algebraic expression $C_{n^*l^*}^2=1/2\pi n^*(2e/n^*)^{2n^*}$. An ADK application are presented in this same book-series by the same authors in \cite{mira2017}. The BSI has been implemented to be used when the ADK model breaks down, the ADK model becomes critic above the threshold
\begin{equation}
E_{cr}\: [V/m]=(\sqrt2-1)  \left( \frac{U_i}{27.2 \mbox{[eV]}} \right)^{3/2} \; \times \; 5.14\cdot10^{14}.
\end{equation} 
We report that the ADK model has also been tested for $m$ values different from zero, but negligible differences have been found for our case scenarios. The BSI effects and uses will be discussed elsewhere.

\section{High quality acceleration experiments at Sparc\_Lab}
In this section we present a possible PWFA Sparc\_Lab case study where the trailing bunch undergoes some acceleration without significant loss of quality. This test case that can be envisioned as a proof of principle experiment aiming in demonstrating the possibility to use PWFA schemes to accelerate bunches with little phase space dilution to finally seed a Free Electron Laser \cite{walker2017,ferrario2018}. This case study assumes both bunches externally generated by a single photo-cathode with a laser modulation, the so called COMB technique. Specifically, in this section we recall the general bunch requirements for this kind of experiment highlighting the choice for our setup. We also discuss the robustness for the presented working point, how bunch transverse mismatching (in size and emmittance) can be partially tolerated without heavily affecting the accelerated quality.

\subsection{Transverse matching condition}
To recall the transverse matching condition, we recall the envelope equation formulation \cite{Reiser,Chao}:
\begin{equation} \label{envelope}
\ddot{\sigma}_x+\frac{\dot{\gamma}}{\gamma} \dot{\sigma}_x + k_{ext}^2 \sigma_x=\frac{\varepsilon^2_{n,rms,x}}{\gamma^2 \sigma^3_x}+\frac{k_{sc}}{\gamma^3 \sigma_x},
\end{equation}
with $\gamma$ is the relativistic factor, $k_{ext}^2=k^2/\gamma m_0 c^2$ with $k$ coefficient of the external linear force ($F_x = k x$), $\varepsilon_{n,rms,x}$ the normalised rms bunch emittance, and with $K_{sc}$  the beam perveance. The derivative along the longitudinal coordinate $z$ has been written as a dot-derivative. Perveance is defined as the ratio of bunch current to the Alfv\'en  current ($I_A=17$ kA) $K_{sc}=2 I_b/I_A$. In this paper we consider emittance-dominated regimes, i.e. perveance contribution can be neglected. In a plasma $k^2_{ext}=e^2 n_p / 2 \epsilon_0 \gamma m_e c^2$ ($e$ elementary charge, $m_e$ electron mass, $c$ speed of light, $n_p$ background number density, $\epsilon_0$ vacuum permittivity and 2 defines the cylindrical symmetry of the system). In the simplified assumption of a constant $\gamma$ the static non oscillating solution is,
\begin{equation} \label{sxmatching}
\sigma_{m} = \sqrt[4]{\frac{2}{\gamma}} \sqrt{\frac{\varepsilon_{n,rms,x}}{k_p}},
\end{equation}
with $k_p$ plasma wavenumber. For the SPARC\_LAB case considering $\gamma=200$ and a background plasma density of the order of $n_p$=\SI{e16}{\cubic\centi\metre}, $\sigma_{m}$ is of the order of a few microns. Assuming a driver with 3 mm-mrad emittance, $\sigma_m$ is \SI{4}{\micro\metre} while \SI{2.3}{\micro\metre} for a trailing bunch with 1 mm-mrad emittance.
The considered driver has a charge of 200 pC, $\sigma_z$=\SI{50}{\micro\metre}, $\sigma_{x,y}$=\SI{4}{\micro\metre}. The trailing bunch has a charge of 10 pC, $\sigma_z$=\SI{10}{\micro\metre}, $\sigma_{x,y}$=\SI{2.35}{\micro\metre}. The distance between the bunches is \SI{167}{\micro\metre}. The driver reduced charge \cite{Rosenzweig2004,Barov1994} is 0.8, corresponding to a weakly non linear regime. ${\tt ALaDyn}$ simulations use a longitudinal resolution of $\Delta z$=\SI{1}{\micro\metre}, and a transverse resolution of $\Delta x=\Delta y$= \SI{0.4}{\micro\metre} for $288\times288\times500$ mesh points; the background is discretised with 8 particle per cell, as for the average number of particles per cell for the bunches. Bunches are initialised in vacuum, the driver travels a distance of 50 $\mu$m while the witness travels a distance of \SI{217}{\micro\metre} before impinging into with the 2 cm long plasma channel.

From the previous considerations we deduce that a key aspect becomes the study of the sensitivity of the identified working point: how much the solution changes by varying one -or more- bunch parameters. We could think this problem as a working point robustness issue. However at present PIC simulations are too expensive to envision a \textit{big-data approach} to the robustness study. Some robustness analysis is being carried out using the hybrid code Architect, but details would be given elsewhere. In the next session we would focus our attention on the specific case on the injection of a driver out of matching.

\subsection{Bunch gaussian shape deterioration over distance: the Shapiro-Wilk test}
\begin{figure}[t]
\begin{center}
\includegraphics{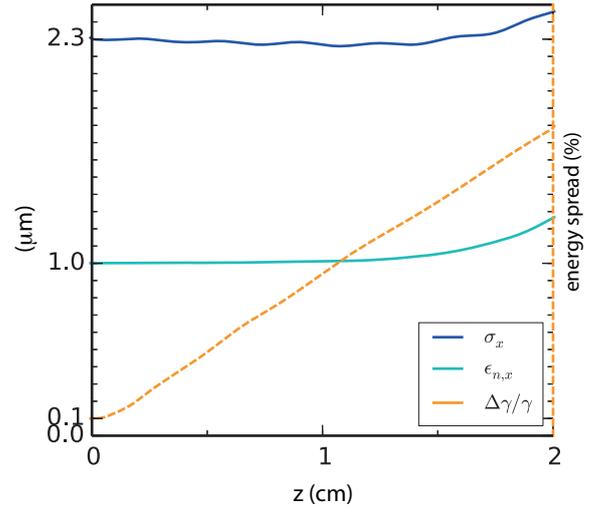}
\caption{Integrated witness parameters: transverse rms-dimension, emittance and energy spread. The energy spread is multiplied by a factor 100 to be displayed on the same graph.}
\label{fig:integratedparameters}
\end{center}
\end{figure}

For the bunch parameters we envision using at our facility, we calculate that the trailing bunch experiences a modest accelerating gradient around 0.8 GV/m. The evolution of integrated parameters, namely transverse rms-dimension, transverse rms-normalised-emittance and energy spread are plotted in Fig.\ref{fig:integratedparameters}. We observe that well controlling the bunch entrance into the plasma, its transverse dimension and its emittance are highly under control and stay stable for the entire accelerating length. The limited trailing bunch charge does not allow a full beam loading, we observe some energy spread growth, but for our accelerating length it remains as low as 1.8\%.

\begin{figure}[t]
\begin{center}
\includegraphics{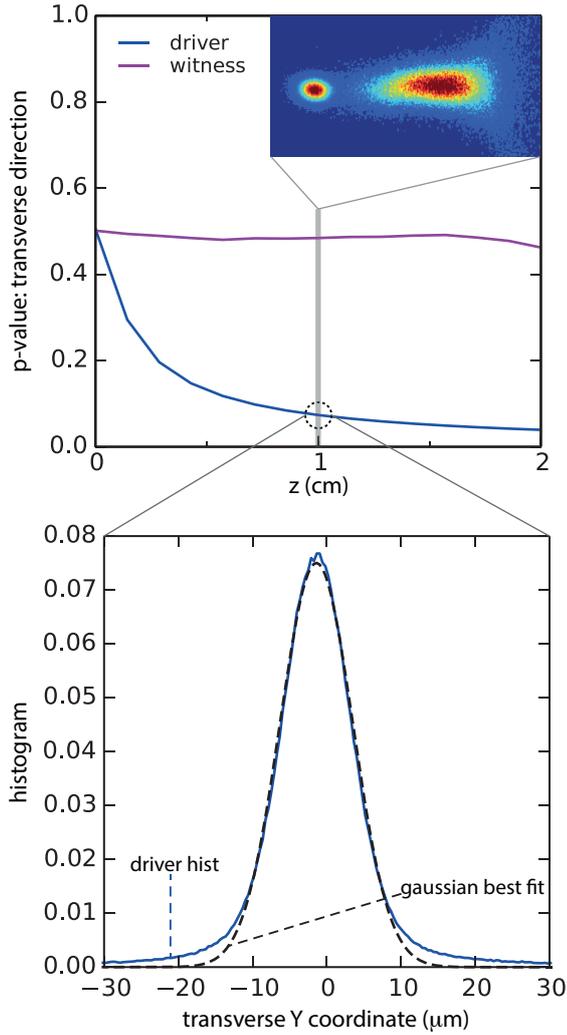}
\caption{Top panel: statistical significance p-value for the Shapiro-Wilk normality-test for the transverse coordinate of both driver and witness. Following the grey guidelines: right upper corner, bunches number density contour plot for the central slice; on the bottom panel, the driver transverse direction coordinate histogram and a comparison with a Gaussian best-fit.}
\label{fig:shapirowilktest}
\end{center}
\end{figure}

The general approach to investigated the bunch quality is via the so called integrated parameters, generally: longitudinal and transverse rms-size, energy spread, mean energy and emittance. In mathematical terms we are studying the moments of the associated probabilistic-density-function distribution. While each moment retains its original meaning, in case the distribution changes over time, the interpretation of the integrated parameters changes. We generally work with particle distribution that have a gaussian distribution in all the three directions. While we wish to check that the trailing bunch keeps its gaussian distribution over time to guarantee an appropriate meaning of each statistical moment, we also want to verify when and how the driver looses such a shape due to the cumbersome nature of the transverse force acting along its length. We propose to combine the classical approach with the Shapiro-Wilk test. We use the test to measure driver and witness degradation from the original shape. A realisation of the significance (p-value) of the test is plotted in Fig.\ref{fig:shapirowilktest}. The trailing bunch retains its gaussian distribution over the entire length (p-value $\gg$ 0.1). More significantly, the test applied to driver suggests that the driver has no longer a transverse Gaussian distribution just after 0.75 cm.

\subsection{Robustness of the working point}
By introducing the working point we have also pointed out that the parameters have been finely tailored. In reality it would be rather difficult to control both bunches simultaneously with the same accuracy. For this reason it is important to evaluate the sensitivity of the witness acceleration by the variation, for example, of the driver rms-dimension or emmittance. In order to simplify our study we focus on two specific cases, the driver is delivered at plasma entrance with a larger dimension than the matched case or the driver is delivered at the plasma entrance with a larger emittance.

\subsubsection{Larger drivers}
The driver is delivered at the plasma entrance out of matching condition, a first case assumes $\sigma_{x,y}=$ \SI{8}{\micro\metre} a second case \SI{16}{\micro\metre}. All cases retain the original 200 pC charge.

\begin{figure}[t]
\begin{center}
\includegraphics{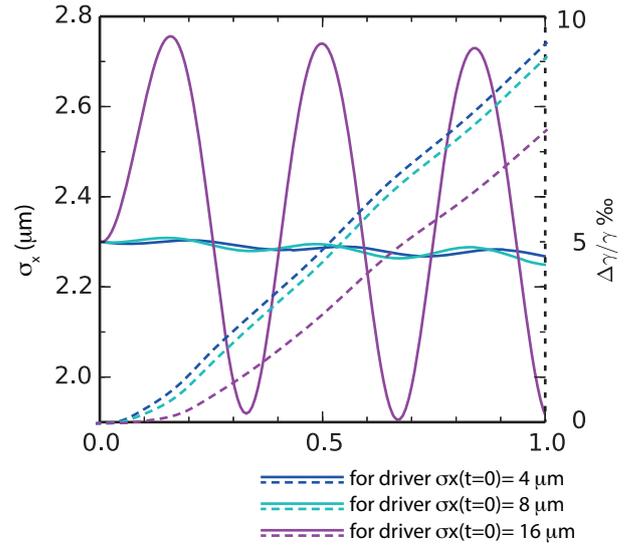}
\caption{Witness main integrated diagnostics for three different driver transverse dimensions: 4 $\mu$m, 8 $\mu$m and 16 $\mu$m. Solid line, trailing bunch $\sigma_x$. Dashed line, trailing bunch energy spread, left $y$ axis.}
\label{fig:witnessrobustness}
\end{center}
\end{figure}

We focus our attention on the witness since our goal is to preserve its overall quality over travelled distance. The degraded cases simulations have been run for 1 cm. We are not interested in the overall evolution but we want to verify whether a strongly mismatched driver seeds any witness instabilities since the very beginning. Fig.\ref{fig:witnessrobustness} reports on the witness transverse dimension and its energy spread. We notice that all witness integrated parameters responds similarly giving no evidence whether our driver is as twice as large as the matching condition, while for the 16 $\mu$m case strong witness oscillations are observed. We notice that these oscillations are relatively very large (20\% of $\sigma_{x,mtc}$) but small in terms of absolute values. While we observe an emittance growth comparable for the three cases, energy spread is lower in the case of the larger driver: the lower charge density induces lower fields that further reduce the energy spread. The driver also in the case of a spread condition is naturally focused by the surrounding plasma that will help increasing the charge density up to the point in which it naturally produces a bubble. The driver stability condition can consequently be over-relaxed once the witness quality is well controlled.

\subsubsection{Drivers with large emittance}
\begin{figure}[t]
\begin{center}
\includegraphics[scale=0.85]{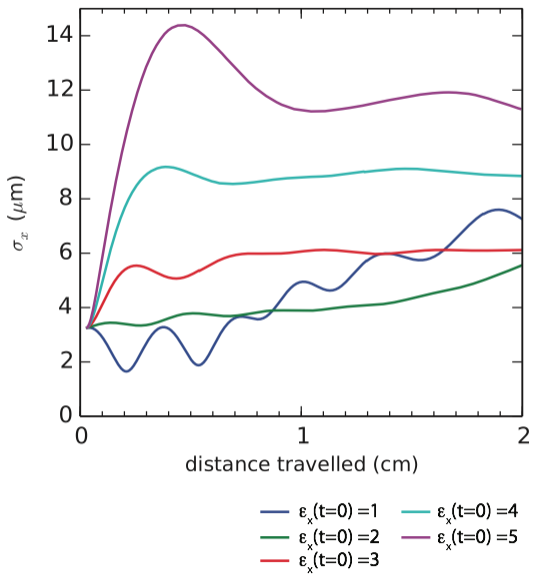}
\caption{Driver central slice evolution for different emittance values, emittance are in unit of mm-mrad.}
\label{fig:emittancepressure}
\end{center}
\end{figure}

We now consider the case for bunch with large emittance. We observe that if the driver is injected with a large emittance it self asses around an equilibrium position. This condition is equivalent of saying that if the bunch is injected smaller than its matched condition, the bunch adiabatically expands up to the point where the emittance is fully balanced by the transverse self generated fields. It can be of very practical use, i.d. if a train of bunch is generated with a COMB technique less constraints are given for the driver(s) generation giving more room for trailing bunch quality optimisation on a pure experimental point. Bunches with larger emittance are also resilient to the hose instability. In order to identify the final matching condition given by the emittance we have balanced the bunch emittance (by interpreting it as pressure) and balancing it with the confining electrostatic pressure. The comparison between this simple balance equation and simulations are reported in Table~1 and in Fig.~\ref{fig:emittancepressure}.

\begin{table}[b]
\caption{Equilibrium transverse size dimension, calculated by Eq.(\ref{epsmatching}) and verified with the particle-in-cell code ${\tt ALaDyn}$}
\begin{center}
\begin{tabular}{|c|c|c|}
\hline
$\varepsilon_x$ (mm-mrad) & $\sigma_{x,\rm{th.}}$ ($\mu$m) & $\sigma_{x,sims}$ ($\mu$m)\\ \hline
1                                                         &   2.0                                             &  2.4        \\
2                                                           &   4.1                                             &  3.9         \\
3                                                           &   6.3                                             &  6.0         \\
4                                                           &   8.4                                             &  8.8         \\
5                                                           &   10.4                                           &  10.0       \\
\hline
\end{tabular}
\end{center}
\label{table1}
\end{table}

\section{Relaxed transverse matching condition for ramped density profile, a simplified analytical case}
\begin{figure}[t]
\begin{center}
\includegraphics{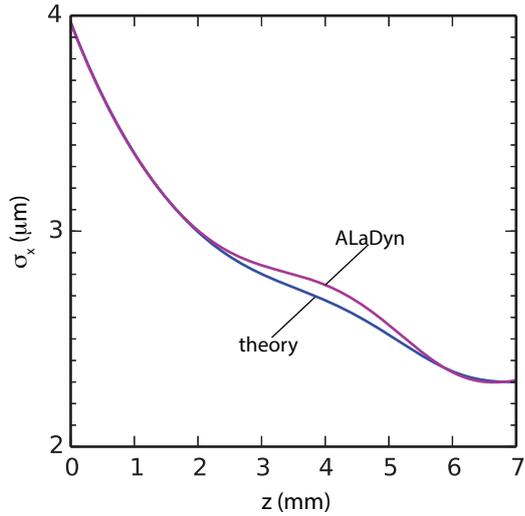}
\caption{Witness envelope evolution inside a linear density ramp. The ramp, 7mm long, is longitudinally linear in density: from vacuum to the nominal density value of $n_p=10^{16}$ cm$^{-3}$. The blue line is the analytical solution, the magenta line is the ${\tt ALaDyn}$ numerically calculated solution.}
\label{rampevolution}
\end{center}
\end{figure}

To extend the calculation for the working point under consideration, we identify a simple rule based on the envelope equation to control bunch injection in the presence of density ramps. The condition investigated throughout the paper of a sharp transition between vacuum and plasma offers a fast and appropriate tool to understand the underlying physics, however in foreseeing a future experiment, density ramps need to be considered. We determine bunch conditions ($\sigma_{x,y}$) at ramp entrance as an envelope equation final value problem. We write $k_{ext}$ of Eq.(\ref{envelope}) as a linear function of density,
\begin{equation}
k_{ext}(z)= z \frac{n_p e^2}{2 \gamma m_e \epsilon_0 c^2}
\end{equation}
where $z$ varies from 0 to $L_{\rm ramp}$, with $L_{\rm ramp}$ ramp length. At $z=0$ with no surrounding plasma the compressing force is null, and it grows linearly with density up to its maximum value where the number density is at nominal value $n_p$. As final conditions, at $z=L_{\rm ramp}$, we require the bunch to be  matched and at waist, i.e. $\sigma_x(t=L_{\rm ramp})=\sigma_{m}$ and $\dot{\sigma}_x(t=L_{\rm ramp})=0$. Eq.(\ref{envelope}) is solved by the coordinate transformation $\zeta=L_{\rm ramp}-z$. Fig.\ref{rampevolution} plots the analytical solution versus the ${\tt ALaDyn}$ numerical solution for $L_{\rm ramp}=$7 mm. We chose a 7 mm long ramp to significantly relax injection conditions and since initial capillary tests suggest a linear ramp profile around 7 mm. The witness is delivered convergent ($\alpha$-Twiss=0.708) with a transverse dimension of almost 4 $\mu$m. Fig.\ref{rampevolution} denotes good agreement between the PIC simulation and the analytical calculation; the small discrepancy around 4 mm is due to driver betatron oscillation that moderately modifies the focusing force. The good agreement especially at the end of the ramp suggests that there is no necessity to include the driver oscillations in the model.

\section{Conclusions}
In this paper we have presented a particle-in-cell numerical investigation for the upcoming PWFA campaign at the SPARC\_LAB test facility. We have investigated a case with two bunches in the weakly non-linear regime. The driver has a charge of 200 pC followed by a witness of 10 pC. The distance between driver and witness is half a plasma wavelength (167 $\mu$m).

We have verified the matching condition formulation in the weakly non linear regime for both bunches. The analytical matching condition applies with no restrictions to the witness, the only precaution being that the witness has to reach the matching condition when the bubble is fully formed around the witness itself. For the case of interest the witness experiences a 0.8 GV/m accelerating gradient keeping both energy spread and emittance below critical thresholds. We have also introduced the usage of the Shapiro-Wilk test to further verify that the witness retains its Gaussian distribution while accelerating, while the driver quickly loses its shape and so its quality.

To verify the robustness of the identified working point we have degraded the driver quality by injecting it into the plasma at large transverse sizes, or eventually by injecting it with larger emittances. For the case at twice the transverse size we notice very small differences; for the case three times larger we reported that strong driver oscillations imply relatively large witness oscillations with consequent witness quality degradation. Large emittances bring the driver to find a natural matching condition at a larger dimension after some expansion, about two betatron oscillations. For our case we observed that a safe threshold injection emittance is around 4 mm-mrad.

We have also briefly reported how plasma ramps can relax matching conditions by injecting the driver at a larger transverse dimension. While we recall a generalised analytical approached based on the envelope equation, we have only shown a specific case based on the presented working point with the simplified case of theoretical linear ramp.

\section{Acknowledgements} \label{sec:Acknowledgements}
The authors acknowledge P. Londrillo for useful discussions. The authors would like to acknowledge INFN-CNAF for providing the computational resources and support required for this work. A. M. and F. M. also acknowledge the CINECA award under the ISCRA initiative, for the availability of high performance computing resources and support. This work was supported by the European Union Horizon 2020 research and innovation programme under grant agreement N. 653782.


%
%
\renewcommand{\baselinestretch}{1.0}

%
%

\end{document}